\documentclass[twocolumn,aps,prl,superscriptaddress,floatfix,noeprint]{revtex4-1}

\usepackage{graphicx,tabularx}
\usepackage{amsmath,amsfonts,amssymb}
\usepackage{bm,dsfont}
\usepackage[colorlinks=true, citecolor=blue, linkcolor=red, urlcolor=blue]{hyperref}
\usepackage[all]{hypcap}
\usepackage{xcolor}
\usepackage[caption=false]{subfig}

\graphicspath{{figures/}}

\DeclareMathOperator{\sgn}{sgn}

\begin{document}

\title{Chiral zigzag modes and flatbands in network models of twisted bilayer graphene}
\author{C. De Beule}
\affiliation{Institute for Mathematical Physics, TU Braunschweig, 38106 Braunschweig, Germany}
\author{F. Dominguez}
\affiliation{Institute for Mathematical Physics, TU Braunschweig, 38106 Braunschweig, Germany}
\author{P. Recher}
\affiliation{Institute for Mathematical Physics, TU Braunschweig, 38106 Braunschweig, Germany}
\affiliation{Laboratory for Emerging Nanometrology, 38106 Braunschweig, Germany}
\date{\today}

\begin{abstract}
We construct a phenomenological scattering theory for the triangular network of valley Hall states that arises in twisted bilayer graphene under interlayer bias. Crucially, our network model includes scattering between different valley Hall states within the same valley and spin. We show that in the absence of forward scattering, symmetries reduce the network model to a single parameter that interpolates between a nested Fermi surface and flatbands, which can be understood in terms of one-dimensional chiral zigzag modes and closed triangular orbits, respectively. We demonstrate how unitarity and symmetry constrain the couplings between zigzag modes, which has important implications on the nature of interference oscillations observed in experiments.
\end{abstract}

\maketitle

In twisted bilayer graphene (TBG) two graphene layers are stacked with a relative twist, leading to a triangular moir\'e pattern of alternating stacking regions which drastically alters the electronic structure \cite{LopesDosSantos2007,Bistritzer2010,Li2010}. In recent years, TBG has garnered immense interest due to the discovery of correlated insulating phases \cite{Kim2017a,Cao2018a}, superconductivity \cite{Cao2018,Yankowitz2019}, ferromagnetism \cite{Sharpe2019}, nematicity \cite{Kerelsky2019,Choi2019}, and strange metals \cite{Cao2019} in magic-angle TBG.

For tiny twist angles ($\theta<1^\circ$) the lattice of TBG relaxes into sharply defined triangular AB/BA stacking domains \cite{Nam2017,Yoo2019,Walet2019}. When a potential bias $\pm U$ is applied between the layers, e.g.\ due to an electric field normal to the layers, a local gap is opened in the AB/BA stacking regions with valley Chern number $N_K = -N_{K'} \approx \pm \sgn(U/\gamma_\perp)$ where $\pm$ corresponds to AB or BA stacking respectively, and $\gamma_\perp$ is the interlayer hopping \cite{Zhang2013}. Consequently, each valley and spin hosts two chiral modes along AB/BA domain walls that propagate in opposite directions for different valleys \cite{Martin2008,Zhang2013,Yin2016}. When the Fermi energy is tuned in the local gap, the low-energy excitations are entirely due to a triangular network of valley Hall states \cite{San-jose2013,Efimkin2018,Huang2018}. Recently, microscopic calculations observed that the network gives rise to one-dimensional (1D) chiral zigzag modes along three independent directions related by $C_3$ rotation symmetry, which leads to a nested Fermi surface with three $C_3$-related nesting vectors \cite{Fleischmann2020,Tsim2020}. However,  current network theories \cite{Efimkin2018} cannot reproduce these results and recent transport experiments that reported interference oscillations are incompatible with decoupled 1D chiral modes \cite{Rickhaus2018,Xu2019}. At the moment, it is unclear how the triplet of 1D chiral zigzag modes arises from the network and how they are coupled.
\begin{figure}
\centering
\includegraphics[width=\linewidth]{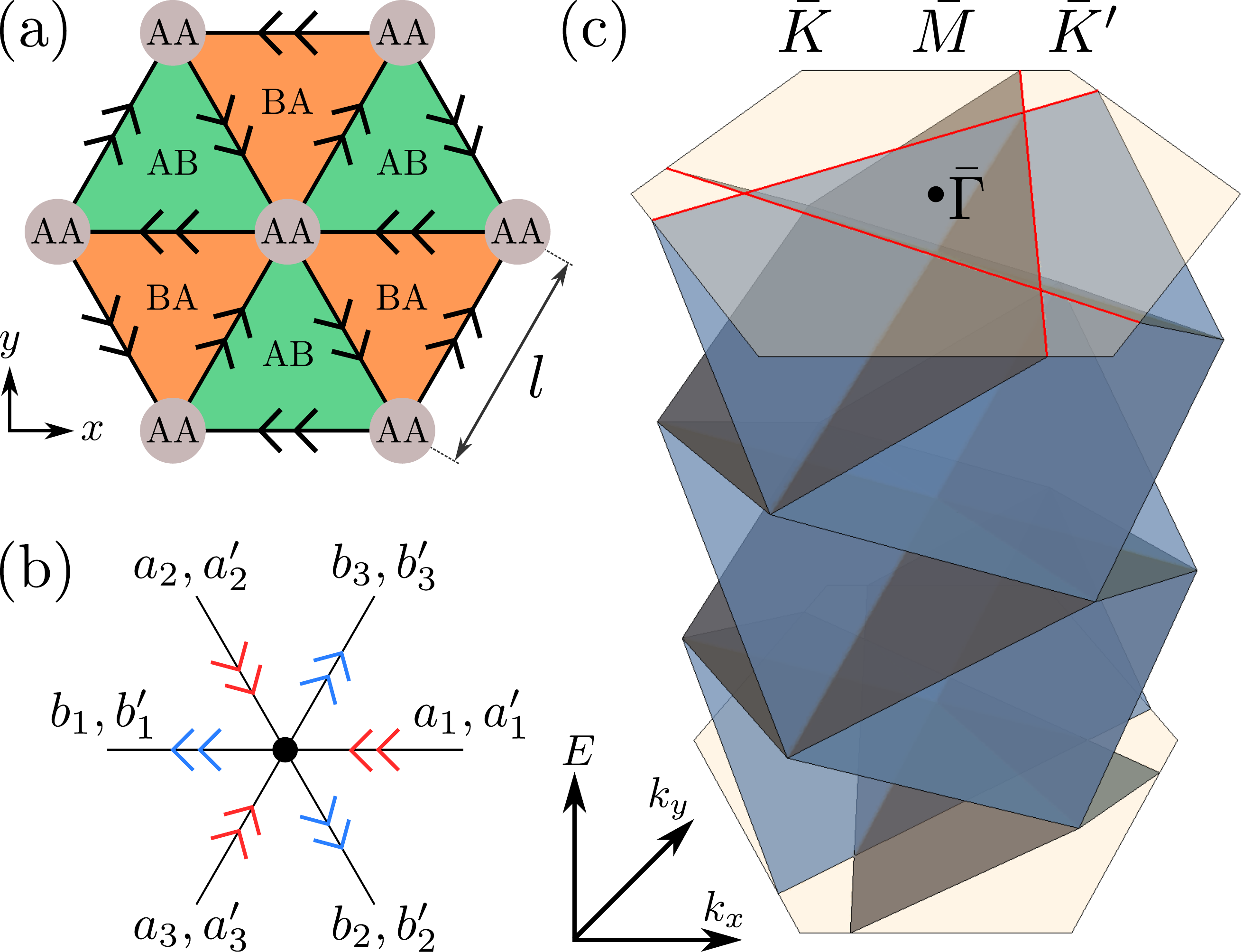}
\caption{(color online) (a) Stacking domains of TBG showing the network of valley Hall states (for a single valley and spin) that emerges upon applying an interlayer bias. AB/BA domain walls and AA regions correspond to the links and scattering nodes of the network model, respectively. (b) Unit cell of the network. (c) Network spectrum for $K$ valley in the moir\'e Brillouin zone (MBZ) for the case with no forward scattering and $\phi=0$ [see Eq.\ \eqref{eq:S0}]. Due to the 1D nature of the network eigenmodes, the Fermi surface (red lines) is nested.}
\label{fig:fig1}
\end{figure}

In this paper, we construct a network model \cite{Chalker1988} for TBG under interlayer bias where the links of the network are given by AB/BA domain walls and the scattering nodes correspond to AA stacking regions, as illustrated in Fig.\ \ref{fig:fig1}(a). While the two valley Hall states do not scatter to each other along links in the absence of disorder, it is not \emph{a priori} clear why they remain decoupled when they reach the AA regions, where the local gap induced by the interlayer bias vanishes. Hence, we allow for scattering between different valley Hall states at the nodes, for a given valley and spin. We do not consider scattering between valleys as the moir\'e pattern varies slowly on the interatomic scale for small twist angles.

Taking into account the symmetries of TBG under interlayer bias and unitarity, we show that in the absence of forward scattering, the network physics is controlled by the phase shift $\phi$ after $120^\circ$ deflections, which tunes the system between 1D chiral zigzag modes and localized modes known as pseudo-Landau levels. We then investigate the robustness of these regimes by including forward scattering, which gives rise to different coupling mechanisms between the zigzag modes. In particular, we find that the robustness of the Fermi surface nesting \cite{Fleischmann2020} can be understood as a consequence of the suppression of forward scattering due to the network geometry.

\emph{Network model} --- We consider a network with two chiral modes along each link which scatter at nodes that form a triangular lattice, as illustrated in Fig.\ \ref{fig:fig1}(a). Each scattering node has six incoming and six outgoing modes as shown in Fig.\ \ref{fig:fig1}(b). We label the nodes by their position vector $\bm r_{ij} = i \bm l_1 + j \bm l_2$ where $\bm l_{1,2} = l (-1/2,\pm \sqrt{3}/2)$ are moir\'e lattice vectors with $l = a/2\sin(\theta/2)$ the moir\'e lattice constant and where $a$ is the lattice constant of graphene. Incoming modes are denoted as $a_{ij} = (a_{1,ij},a_{2,ij},a_{3,ij})$ and $a_{ij}'$ for the two chiral channels, while outgoing modes are denoted as $b_{ij}$ and $b_{ij}'$, such that $(b,b')^t=\mathcal S(a,a')^t$ with $\mathcal S$ the $S$-matrix relating incoming to outgoing modes.

To constrain the $S$-matrix, we take into account the symmetries of TBG under interlayer bias. At small twist angles, the symmetries of TBG become independent of the twist center \cite{Po2018,Zou2018} so that we do not have to consider a specific lattice realization. Symmetries that preserve the valley are given by $C_3$ and $C_2 T$, where $T$ is (spinless) time-reversal symmetry and $C_3$ and $C_2$ are rotations by $2\pi/3$ and $\pi$ about the $z$-axis with respect to the center of an AA region, respectively. Note that $C_2$ exchanges both the A and B sublattices and valleys. These symmetries impose the following conditions on the $S$-matrix \footnote{See supplemental material [url to be added].}:
\begin{alignat}{3}
& C_3: \quad && \mathcal S = C_3 \mathcal S C_3^{-1}, \\
& C_2: \quad && \mathcal S_{K'} = \mathcal S_{K}, \\
& T: \quad && \mathcal S_{K'} = ( \mathcal S_{K} )^t, \\
& C_2 T: \quad && \mathcal S = \mathcal S^t,
\end{alignat}
where $C_3$ corresponds to a cyclic permutation of the incoming modes $(a_1,a_1') \rightarrow (a_2,a_2') \rightarrow (a_3,a_3') \rightarrow (a_1,a_1')$ and similar for outgoing modes.

To proceed, we first neglect forward scattering, which is a good starting point as the wave-function overlap between incoming and outgoing modes is larger for deflections than for forward scattering, simply due to the geometry of the triangular network \cite{Qiao2014a}. In contrast to previous network models for TBG under interlayer bias \cite{Efimkin2018}, we take into account scattering at the nodes (AA regions) between the two chiral modes belonging to the same valley and spin. It can be shown that up to a unitary transformation \cite{Note1}, the most general $S$-matrix obeying $C_3$ and $C_2T$ symmetry in the absence of forward scattering is given by
\begin{equation} \label{eq:S0}
\mathcal S = \frac{e^{i\varphi}}{2} \begin{pmatrix}
S_{\phi,\phi} & S_{0,\pi} \\
S_{\pi,0} & -S_{-\phi,-\phi}
\end{pmatrix}, 
\end{equation}
where
\begin{equation}
S_{\vartheta,\psi} =
\begin{pmatrix}
0 & e^{i\vartheta} & e^{i\psi} \\
e^{i\psi} & 0 & e^{i\vartheta} \\
e^{i\vartheta} & e^{i\psi} & 0
\end{pmatrix},
\end{equation}
with $\varphi$ real and $0 \leq \phi \leq \pi/2$. Using Bloch's theorem, we relate the incoming modes to the outgoing modes of the same node, $(a, a')^t= e^{iEl/\hbar v} \left[ \mathds 1_2 \otimes \mathcal M(\bm k) \right](b,b')^t$ where $e^{iEl/\hbar v}$ is the dynamical phase accumulated along a link with $v$ the velocity of the chiral modes, which we assume is equal for the two valley Hall states, and $\mathcal M(\bm k) = \textrm{diag} \left( e^{ik_3}, e^{ik_1}, e^{ik_2} \right)$ with $k_j = \bm k \cdot \bm l_j$ ($j=1,2,3$) and $\bm l_3 = -(\bm l_1+\bm l_2)$. The network energy bands are then found from $\det \left( \mathds 1_6 -  e^{iEl/\hbar v} \left[ \mathds 1_2 \otimes \mathcal M(\bm k) \right] \mathcal S \right) =0$ \cite{Efimkin2018,Pal2019}.

The phase shift $\phi$ in Eq.\ \eqref{eq:S0} acquired after deflections determines the interference between the network modes and should depend on microscopic parameters such as the Fermi energy, interlayer bias, twist angle, etc.. However, here we treat $\phi$ as a phenomenological parameter. In particular, for $\phi = 0$, the network spectrum becomes
\begin{equation} \label{eq:spectrum1}
E_{j,n}(\bm k) = \frac{\hbar v}{2l} \left( 2 \pi n - 2\varphi + k_j \right),
\end{equation}
where $n\in \mathbb Z$ and which is shown in Fig.\ \ref{fig:fig1}(c). The network spectrum is periodic in energy, in this case with period $\pi\hbar v/l$, and $E_{i,n}(\bm k + \bm g) = E_{i,n+m}(\bm k)$ with $\bm g$ a moir\'e reciprocal lattice vector and $m$ an integer.
\begin{figure}
\centering
\includegraphics[width=\linewidth]{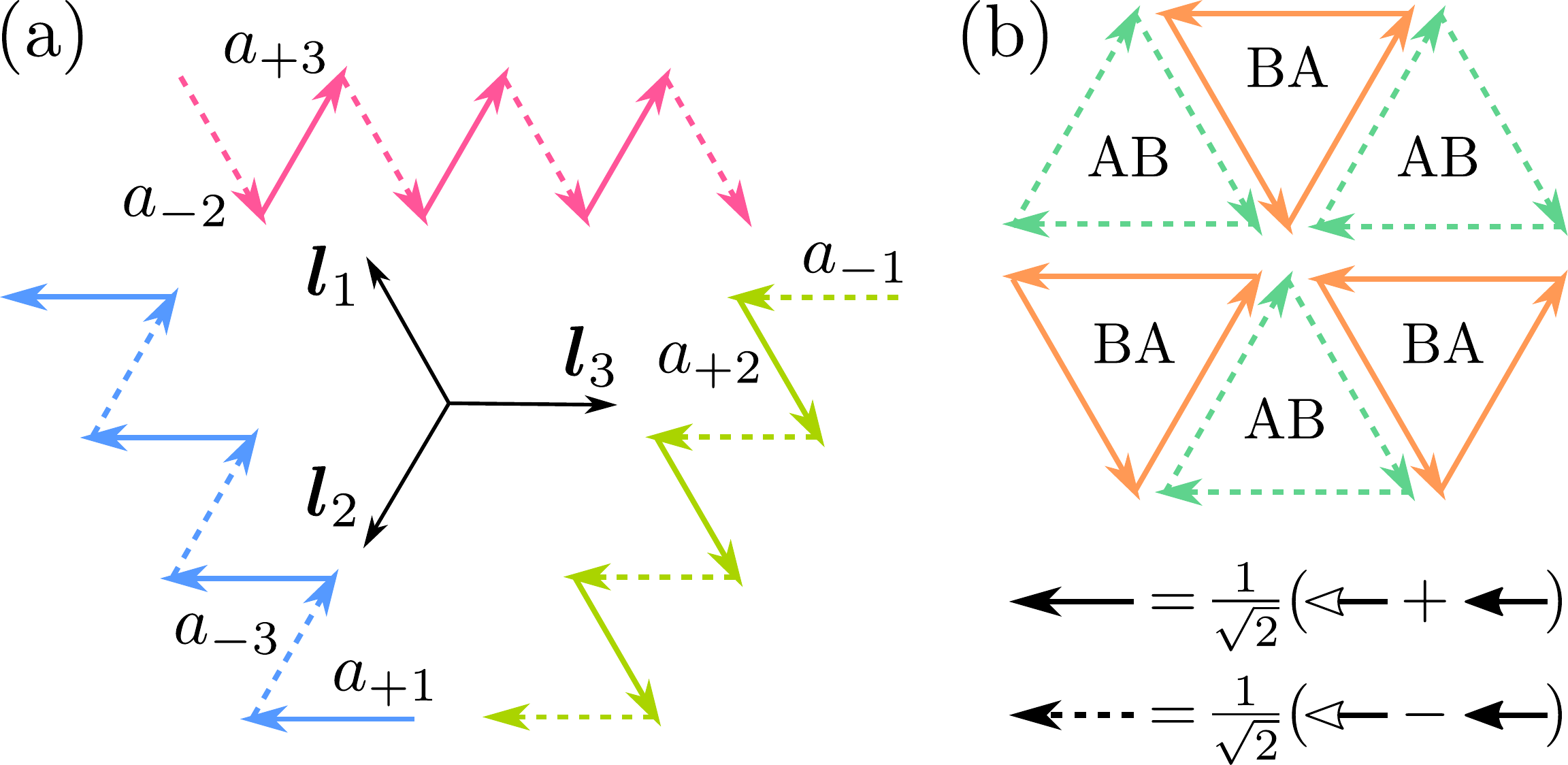}
\caption{(color online) (a) Triplet of 1D chiral zigzag modes ($\phi = 0$) along directions $\bm l_j$ ($j=1,2,3$) with $\bm l_3 = -(\bm l_1+\bm l_2)$, where solid (dashed) lines are (anti)symmetric superpositions of valley Hall states along the same link $a_\pm = (a \pm a')/\sqrt{2}$. (b) Localized network ($\phi = \pi/2$) and illustration of the superpositions of $a$ and $a'$ (solid and open arrowheads).}
\label{fig:fig2}
\end{figure}
To gain some insight, we perform a unitary transformation $U = \left[ \mathds 1_6 + i\sigma_y e^{i\phi\sigma_z} \otimes \mathds 1_3 \right]/\sqrt{2}$ on the scattering matrix, which corresponds to changing the original basis $a,a'$ to a basis of symmetric and antisymmetric superpositions (SAS) of valley Hall states on the same link $a_\pm = (a\pm a')/\sqrt{2}$ and similar for outgoing modes. In the new basis, there are only interchannel deflections \cite{Note1} that proceed in clockwise (counterclockwise) fashion for $a_+$ ($a_-$) as illustrated in Fig.\ \ref{fig:fig2}(a), giving rise to three independent 1D chiral zigzag channels \cite{Tsim2020}. Owing to their linear dispersion, the density of states of the zigzag modes is constant and given by $2\sqrt{3}/\pi \hbar v l$, such that each band with width $\pi \hbar v/l$ hosts one electron per moir\'e unit cell (for each valley and spin). On the other hand, the opposite limit $\phi = \pi/2$ results in three doubly-degenerate flatbands per energy period $2\pi\hbar v/l$, given by $(j=0,1,2)$
\begin{equation}
E_{j,n}(\bm k) = \frac{\hbar v}{l} \left( 2\pi n - \varphi + \frac{\pi}{6} + \frac{2\pi}{3} j \right),
\end{equation}
that we identify with pseudo-Landau levels \cite{Ramires2018,Tsim2020}. In the SAS basis, there are now only intrachannel deflections, such that $a_+$ ($a_-$) modes perform counterclockwise (clockwise) orbits around BA (AB) domains as shown in Fig.\ \ref{fig:fig2}(b). Hence, the network modes are localized, leading to flatbands. Because the orbits consist of superpositions of two chiral modes with different momenta, we expect a non-trivial standing wave pattern \cite{Ramires2018}.

We thus find that in the absence of forward scattering, the phase shift $\phi$ tunes the network between chiral zigzag modes $(\phi=0)$ and flatbands $(\phi=\pi/2)$. For intermediate $\phi$, there is a crossover where zigzag modes are coupled to the localized modes.
\begin{figure}
\centering
\includegraphics[width=\linewidth]{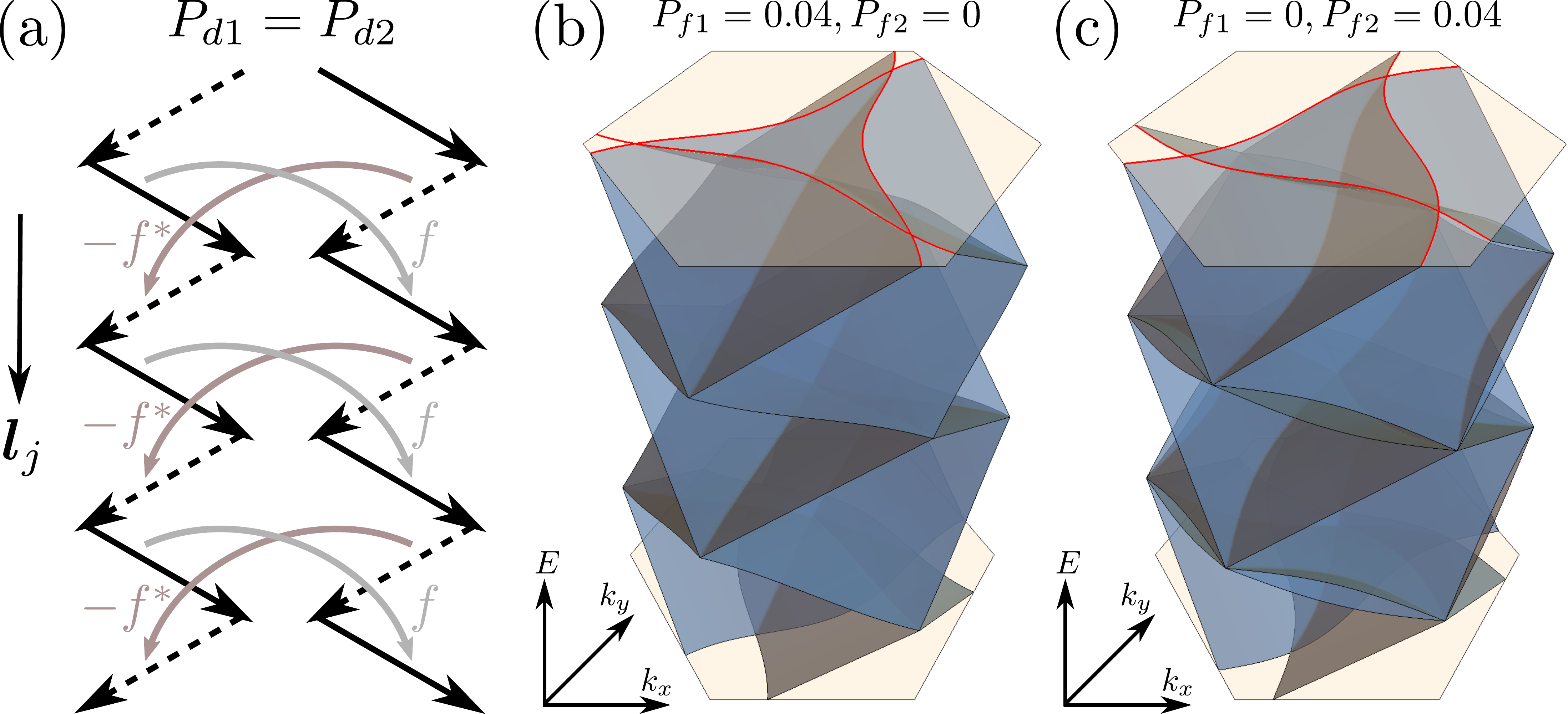}
\caption{Effects of forward scattering on the zigzag network with $P_{d1}=P_{d2}$ and $\phi=0$. (a) Coupling mechanism between parallel channels along $\bm l_j$ ($j=1,2,3$). (b) Network spectrum for $P_{f1} = 0.04$ and $P_{f2} = 0$ and (c) $P_{f1} = 0$ and $P_{f2} = 0.04$.}
\label{fig:fig3}
\end{figure}

\emph{Coupling of zigzag modes ---} We now explore the effect of forward scattering between valley Hall states. For concreteness, we set $\varphi = 0$ and we include intra- and interchannel forward scattering with probabilities $P_{f1}$ and $P_{f2}$. We first consider the case $\phi=0$ and allow for different intra- and interchannel deflection probabilities $P_{d1}$ and $P_{d2}$. Here, we assume that the intrachannel probabilities $P_{f1}$ and $P_{d1}$ are equal for the two valley Hall states. In this case, current conservation requires $2(P_{d1} + P_{d2}) + P_{f1} + P_{f2} = 1$, and we find \cite{Note1}
\begin{equation} \label{eq:S1}
U^\dag \mathcal S U = \begin{pmatrix} 
f \mathds 1_3 & S_0 \\
S_0^t & -f^* \mathds 1_3
\end{pmatrix},
\end{equation}
where $f = \sqrt{P_{f2}} + i \sqrt{P_{f1}} \,  \sin \chi$,
\begin{equation} \label{eq:S2}
S_0 = \frac{-\delta_+ \delta_-}{\delta_+ + \delta_-} \, \mathds 1_3 + \begin{pmatrix}
0 & \delta_+ & \delta_- \\
\delta_- & 0 & \delta_+ \\
\delta_+ & \delta_- & 0
\end{pmatrix},
\end{equation}
with $\delta_\pm = \sqrt{P_{d1}} \pm \sqrt{P_{d2}}$, and $\cos\chi = \left( P_{d2} - P_{d1} \right) / 2 \sqrt{P_{f1} P_{d1}}$. Note that $\mathcal S$ is only well-defined if $\chi$ is real, i.e.\ $2 \sqrt{P_{f1} P_{d1}} \geq | P_{d2}-P_{d1} |$.
\begin{figure}
\centering
\includegraphics[width=\linewidth]{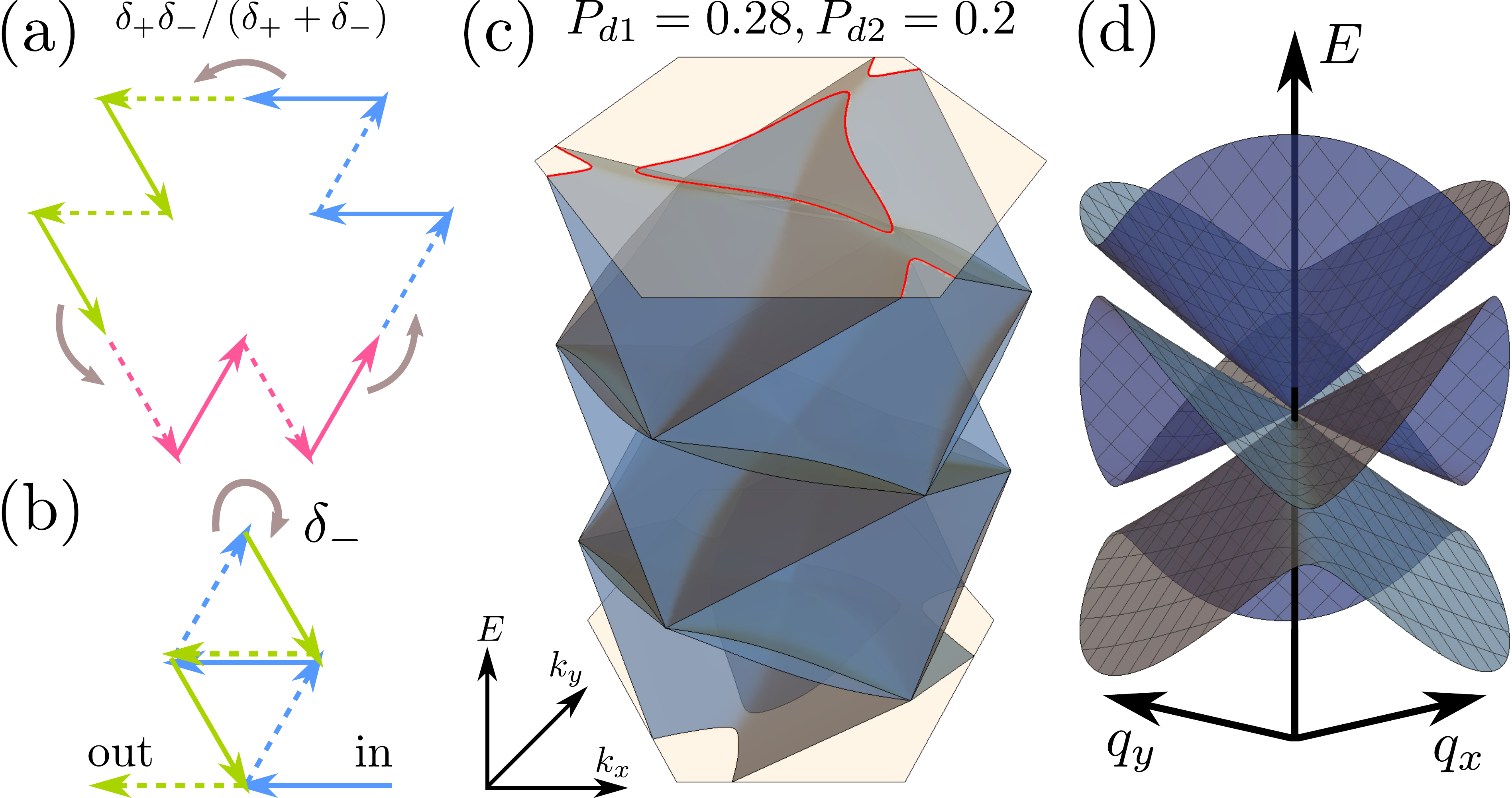}
\caption{(color online) (a,b) Couplings between different zigzag modes which arise due to $P_{d1} \neq P_{d2}$ for $\phi=0$. (c) Network spectrum for $P_{f1} = P_{f2} = 0.02$, $P_{d1} = 0.28$, and $P_{d2} = 0.2$. (d) Zoom of (c) near $\bar K$ with $\bm q = \bm k - \bar K$.}
\label{fig:fig4}
\end{figure}

When $P_{d1}=P_{d2}$, for which $\delta_-=0$ and $\chi=\pi/2$, we find that only parallel zigzag channels are coupled due to intrachannel forward scattering in the SAS basis with probability $|f|^2$, which is illustrated in Fig.\ \ref{fig:fig3}(a). The network spectrum is now given by $(j=1,2,3)$
\begin{equation} \label{eq:spectrum2}
\begin{aligned}
& E_{j,\pm,n}(\bm k) = \\
& \frac{\hbar v}{l} \left[ 2 \pi n - i \log \frac{1}{2} \left( F_j(\bm k) \pm \sqrt{4e^{i k_j} + F_j(\bm k)^2} \right) \right],
\end{aligned}
\end{equation}
where $F_j(\bm k) = f^* e^{-ik_{j+1}} - f e^{-ik_{j+2}}$ with $j$ defined cyclically and which is shown in Figs.\ \ref{fig:fig3}(b) and (c). We see that coupling between parallel zigzag channels warps the Fermi surface, in a manner depending on the type of forward scattering. For $P_{f2}=0$, the bands are symmetric about $k_y$ as in this case $F_1(k_x,-k_y)=F_2(k_x,k_y)$ and $F_3(k_x,-k_y)=F_3(k_x,k_y)$. Furthermore, states at the $\bar \Gamma$ and $\pm\bar K$ points in the moir\'e Brillouin zone (MBZ) remain triply degenerate, but are shifted as $F_j(\bar \Gamma) = -2i\sqrt{P_{f1}}$ and $F_j(\pm \bar K) = \pm 2 e^{\pm i\pi/6} \sqrt{P_{f1}}$ for all $j$.

\begin{figure}
\centering
\includegraphics[width=\linewidth]{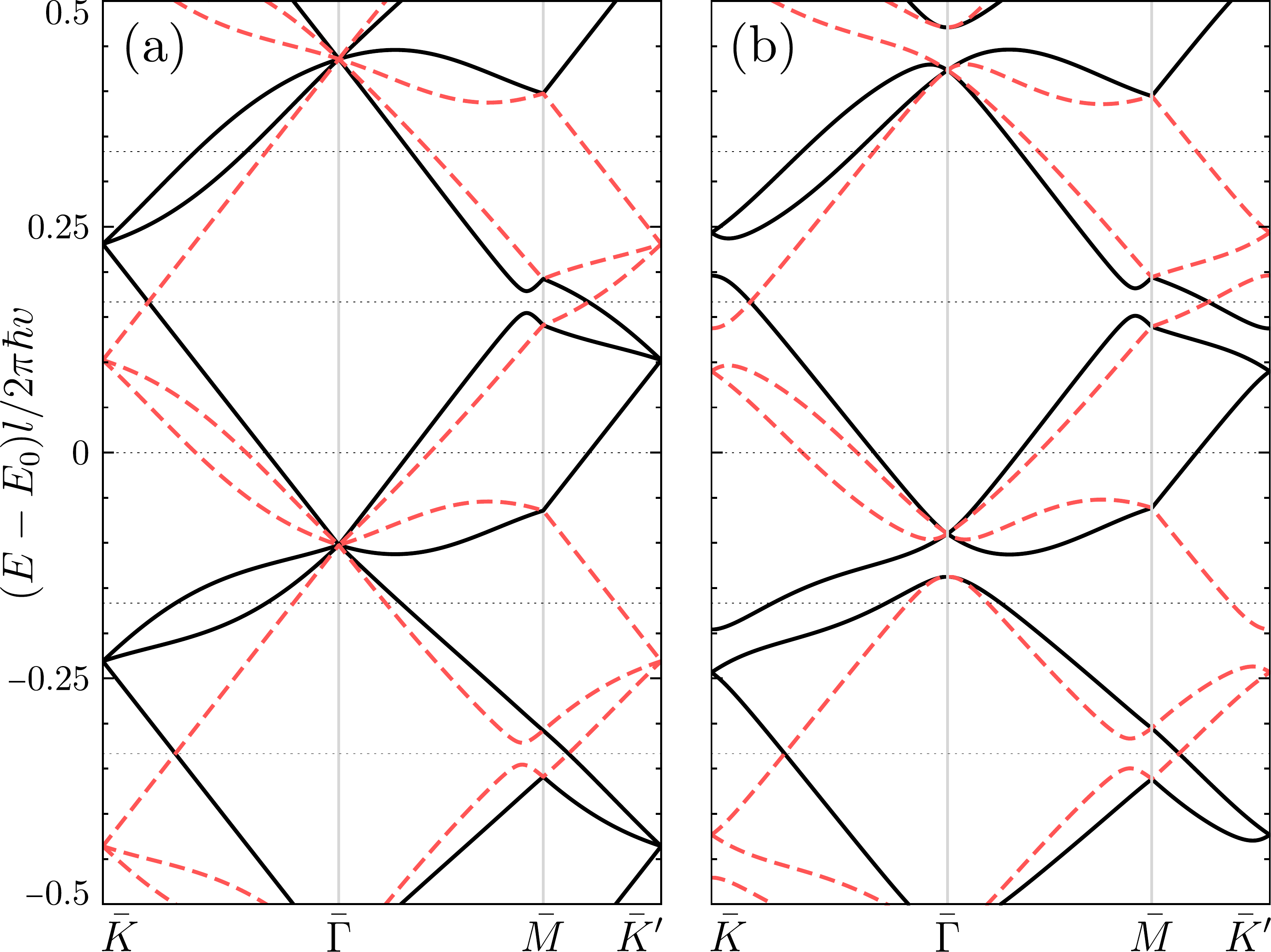}
\caption{Network bands for coupled zigzag modes along high-symmetry lines in the MBZ for $K$ (solid) and $K'$ (dashed) with $E_0 = \pi \hbar v/6l$. (a) $P_{f1}=P_{f2}=0.02$, $P_{d1} =0.28$, $P_{d2}=0.2$, and $\phi=0$. (b) $P_{d1} = P_{d2}$ and $\phi=0.2$.}
\label{fig:fig5}
\end{figure}

\begin{figure}
\centering
\includegraphics[width=\linewidth]{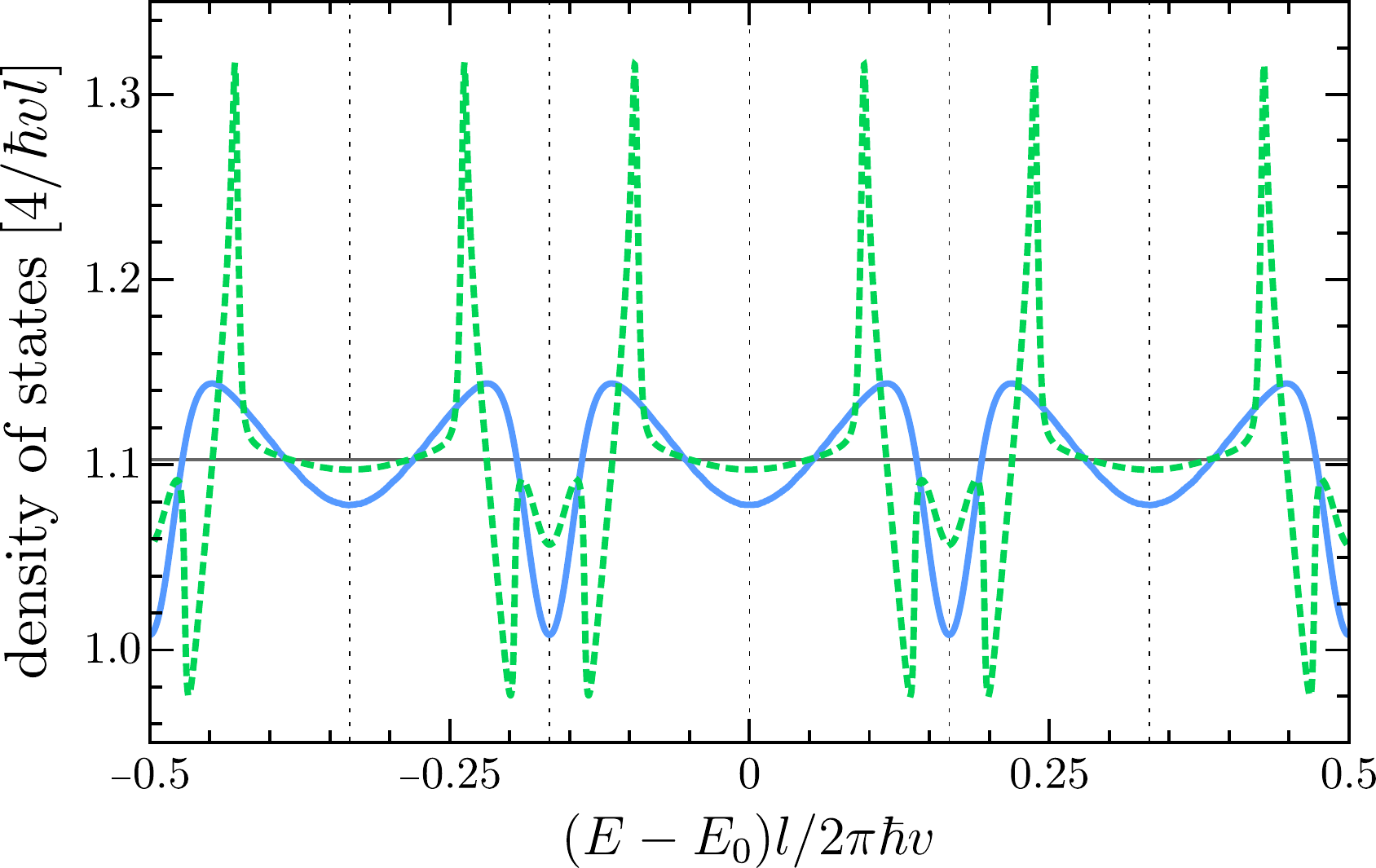}
\caption{Density of states (DOS) over one energy period $2\pi \hbar v/l$ with $P_{f1}=P_{f2}=0.02$ for  (solid) $P_{d1} = 0.28$, $P_{d2}=0.2$, and $\phi=0$ [see Fig.\ \ref{fig:fig5}(a)] and (dashed) $P_{d1} = P_{d2}$ and $\phi=0.2$ [see Fig.\ \ref{fig:fig5}(b)]. The horizontal line gives the constant DOS for decoupled zigzag modes, i.e.\ $8 \sqrt{3}/\pi \hbar v l$.}
\label{fig:fig6}
\end{figure}

In general, Eq.\ \eqref{eq:S2} tells us that the three zigzag channels are coupled through two processes, illustrated in Figs.\ \ref{fig:fig4}(a) and (b). One process is due to interchannel forward scattering in the SAS basis with probability $\left[ \delta_+ \delta_- / \left( \delta_+ + \delta_- \right) \right]^2$ (see Fig.\ \ref{fig:fig4}(a)), while the other is due to clockwise (counterclockwise) deflections from antisymmetric (symmetric) to symmetric (antisymmetric) superpositions with probability $\delta_-^2$ (see Fig.\ \ref{fig:fig4}(b)). Both processes lead to anti-crossings in the network spectrum as can be seen in Fig.\ \ref{fig:fig4}(c). In this case, there are no analytical solutions. Nevertheless, we find that network bands belonging to different zigzag modes hybridize, except at the $\bar \Gamma$ and $\pm \bar K$ points in the MBZ, as shown in Fig.\ \ref{fig:fig4}(d) and Fig.\ \ref{fig:fig5}(a). These crossings give rise to maxima in the density of states (DOS) shown in Fig.\ \ref{fig:fig6}. Minima in the DOS occur at energies in between the nodes where the anti-crossings are largest.

The zigzag modes are also coupled if we allow for a phase shift $\phi > 0$. Since we already discussed the effects of $P_{d1} \neq P_{d2}$ for $\phi=0$, we set $P_{d1} = P_{d2} \equiv P_d$ in this case. The $S$-matrix becomes,
\begin{equation} \label{eq:S3}
U^\dag \mathcal S U = \begin{pmatrix} 
S_1 & S_2^\dag \\
S_2 & -S_1^\dag
\end{pmatrix},
\end{equation}
with $S_1 = f \cos \phi \, \mathds 1_3 + i \sin \phi \, S_0$ and $S_2 = ie^{i\phi} f \sin \phi \, \mathds 1_3 + e^{i\phi} \cos \phi \, S_0$. We find that the scattering amplitude between parallel channels is reduced by a factor $\cos \phi$ and that zigzag modes along different directions are coupled by a similar process as shown in Fig.\ \ref{fig:fig4}(a) but instead with amplitude $ie^{i\phi} f \sin \phi$. Additionally, the zigzag modes are coupled via the localized modes (see Fig.\ \ref{fig:fig2}(b)) through deflections with amplitude $2i\sqrt{P_d} \, \sin \phi$. The corresponding network spectrum and density of states is shown in Fig.\ \ref{fig:fig5}(b) and Fig.\ \ref{fig:fig6}, respectively. Contrary to the previous case, the triple degeneracy at $\bar \Gamma$ and $\pm \bar K$ is reduced to a single crossing protected by $C_3$ and $C_2 T$.

With these results, we can understand the robustness of the Fermi surface nesting in the zigzag regime as reported in Ref.\ \onlinecite{Fleischmann2020}. Due to the geometry of the triangular network, forward scattering is suppressed as the wave-function overlap is smaller \cite{Qiao2014a}. In this case, unitarity automatically enforces $\delta_- \approx 0$ through the condition $2 \sqrt{P_{f1} P_{d1}} \geq | P_{d2}-P_{d1} |$ and therefore coupling between different zigzag channels is suppressed. On the other hand, Eq.\ \eqref{eq:S3} shows that the flatbands ($\phi=\pi/2$) are not robust against forward scattering. This is also observed in band structure calculations as the pseudo-Landau levels disappear when lattice relaxation is taking into account, which leads to sharper domain walls and more forward scattering \cite{Tsim2020}.

\emph{Conclusions} --- We have constructed a phenomenological scattering theory for the triangular network of valley Hall states that arises at low-energies in twisted bilayer graphene under interlayer bias. Our model is based solely on the symmetries of twisted bilayer graphene and unitarity of the $S$-matrix. In the absence of forward scattering, we showed that the network model depends only on the phase picked up after intrachannel deflections, which tunes the system between a nested Fermi surface and pseudo-Landau levels. In this sense, we give a unified explanation of these two phenomena, both arising from the network, in terms of one-dimensional chiral zigzag modes and closed triangular orbits. Moreover, external control over this phase shift would allow one to tailor the properties of the network. We have also explored the effect of forward scattering between valley Hall states on the chiral zigzag modes. In particular, we have shown that the robustness of the nesting arises due to the geometry of the triangular network, which suppresses forward scattering and conspires with unitarity such that zigzag channels propagating in different directions remain largely decoupled. Finally, we addressed different coupling mechanisms between zigzag modes, which have important implications on electronic transport in the network, especially the nature of interference oscillations observed in recent experiments \cite{Xu2019,Rickhaus2018}. The network model has a rich phenomenology but is simple enough at the same time to allow for qualitative predictions.

\begin{acknowledgments}
\emph{Acknowledgments} --- F.D.\ and P.R.\ gratefully acknowledge funding by the Deutsche Forschungsgemeinschaft (DFG, German Research Foundation) within the framework of Germany's Excellence Strategy -- EXC-2123 QuantumFrontiers -- 390837967.
\end{acknowledgments}

\bibliography{references.bib}


\pagebreak
\onecolumngrid

\begin{center}
\textbf{\large Supplemental Material}
\end{center}

\setcounter{equation}{0}
\setcounter{figure}{0}
\setcounter{table}{0}
\setcounter{page}{1}

\renewcommand{\thepage}{S\arabic{page}}
\renewcommand{\thesection}{S\arabic{section}}  
\renewcommand{\thetable}{S\arabic{table}}  
\renewcommand{\thefigure}{S\arabic{figure}}

\section{S1.\quad Symmetry constraints on the S-matrix}

First, we consider $C_3$ rotation symmetry which preserves the valley. We find that
\begin{equation}
\mathcal S = C_3 \mathcal S C_3^{-1},
\end{equation}
where $C_3$ is a cyclic permutation $(a_1,a_1') \rightarrow (a_2,a_2') \rightarrow (a_3,a_3') \rightarrow (a_1,a_1')$ of the incoming modes which are defined in Fig.\ {\color{red} 1}(b) of the main text, and similar for outgoing modes. Next, we discuss the effect of $C_2$ rotation symmetry and time-reversal symmetry $T$. As these symmetries do not conserve the valley, we need to consider both valleys:
\begin{equation}
\begin{pmatrix} b_K \\ b_{K'} \end{pmatrix} = \begin{pmatrix} \mathcal S_K & 0 \\ 0 & \mathcal S_{K'} \end{pmatrix} \begin{pmatrix} a_K \\ a_{K'} \end{pmatrix}.
\end{equation}
Under $C_2$ rotation symmetry, we have
\begin{equation}
\begin{pmatrix} b_{K'} \\ b_K \end{pmatrix} = \begin{pmatrix} \mathcal S_K & 0 \\ 0 & \mathcal S_{K'} \end{pmatrix} \begin{pmatrix} a_{K'} \\ a_K \end{pmatrix},
\end{equation}
such that $\mathcal S_{K'} = \mathcal S_K$. On the other hand, under time-reversal symmetry we have
\begin{equation}
\begin{pmatrix} a_{K'}^* \\ a_K^* \end{pmatrix} = \begin{pmatrix} \mathcal S_K & 0 \\ 0 & \mathcal S_{K'} \end{pmatrix} \begin{pmatrix} b_{K'}^* \\ b_K^* \end{pmatrix},
\end{equation}
such that $\mathcal S_{K'} = (\mathcal S_K)^t$. Hence, the combination $C_2T$ enforces $\mathcal S_K = (\mathcal S_K)^t$.

\section{S2.\quad S-matrix without forward scattering}

\begin{figure}
\centering
\includegraphics[width=\linewidth]{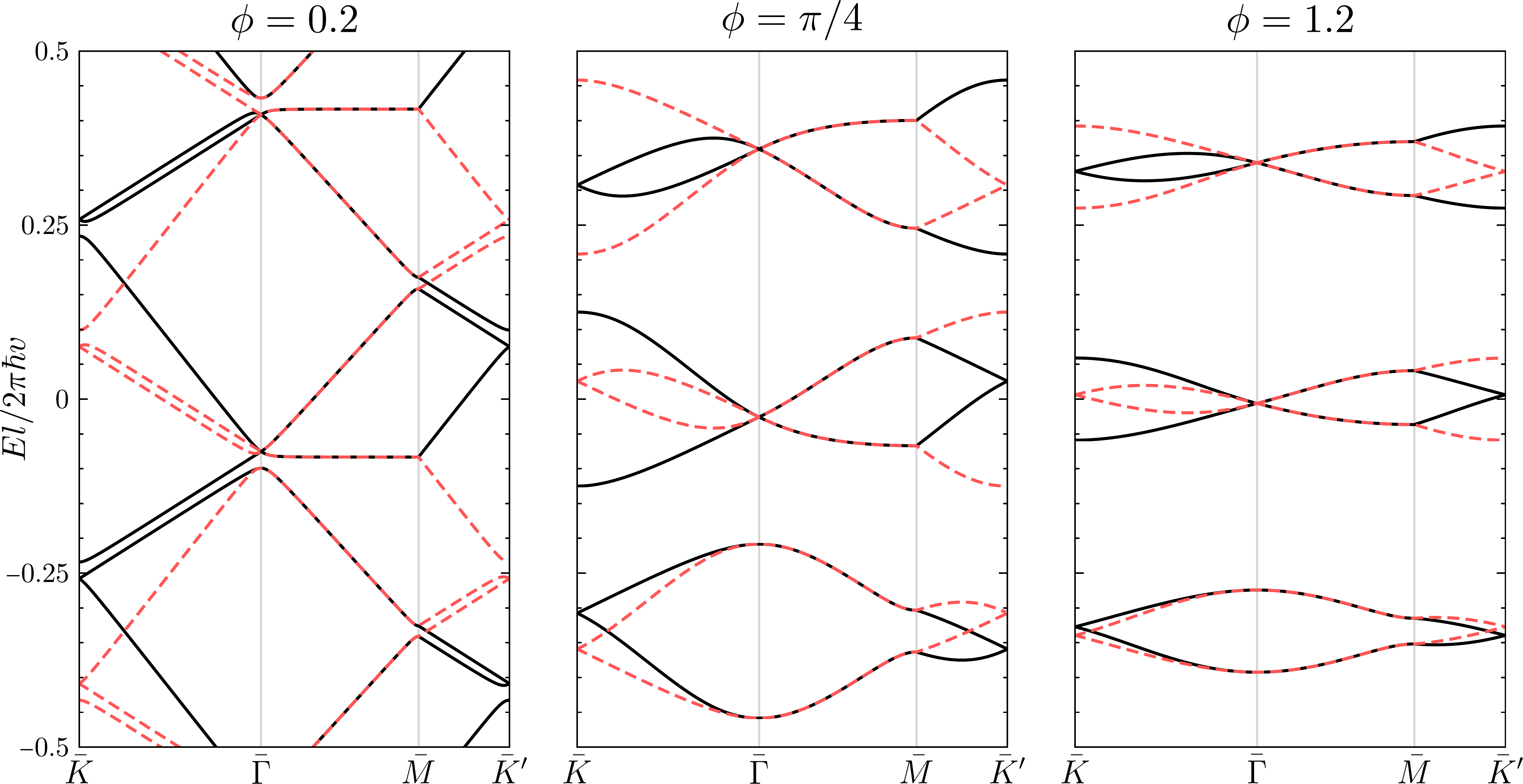}
\caption{Network spectrum in the absence of forward scattering over one energy period $2\pi \hbar v/l$ along high-symmetry lines of the moir\'e Brillouin zone with $\varphi=-\pi \hbar v/6l$. Solid (dashed) curves correspond to the $K$ ($K'$) valley.}
\label{fig:sfig1}
\end{figure}

The $S$-matrix relates valley Hall states that propagate along AB/BA domain walls at the scattering nodes (AA regions) such that $(b,b')^t = \mathcal S (a, a')^t$ with $a,a'$ six incoming modes and $b,b'$ six outgoing modes where the prime distinguishes the two valley Hall states as illustrated in Fig.\ {\color{red} 1}(b) of the main text. In the absence of forward scattering, we find that the most general $S$-matrix consistent with unitarity and $C_3$ and $C_2T$ symmetry is given by
\begin{equation} \label{eq:Smat1}
\mathcal S = e^{i\varphi} \begin{pmatrix}
0 & e^{i \phi} \sqrt{P_{d1}} & e^{i \phi} \sqrt{P_{d1}} & 0 & \sqrt{P_{d2R}} & -\sqrt{P_{d2L}} \\
e^{i \phi} \sqrt{P_{d1}} & 0 & e^{i \phi} \sqrt{P_{d1}} & -\sqrt{P_{d2L}} & 0 & \sqrt{P_{d2R}} \\
e^{i \phi} \sqrt{P_{d1}} & e^{i \phi} \sqrt{P_{d1}} & 0 & \sqrt{P_{d2R}} & -\sqrt{P_{d2L}} & 0 \\
0 & -\sqrt{P_{d2L}} & \sqrt{P_{d2R}} & 0 & -e^{-i\phi} \sqrt{P_{d1}} & -e^{-i\phi} \sqrt{P_{d1}} \\
\sqrt{P_{d2R}} & 0 & -\sqrt{P_{d2L}} & -e^{-i\phi} \sqrt{P_{d1}} & 0 & -e^{-i\phi} \sqrt{P_{d1}} \\
-\sqrt{P_{d2L}} & \sqrt{P_{d2R}} & 0 & -e^{-i\phi} \sqrt{P_{d1}} &-e^{-i\phi} \sqrt{P_{d1}} & 0
\end{pmatrix},
\end{equation}
with $\varphi$ and $\phi$ real phases, and with the conditions $2P_{d1}+P_{d2R}+P_{d2L}=1$ and $P_{d1}=\sqrt{P_{d2R} P_{d2L}}$ which has two solutions. Either all probabilities are nonzero with $0<P_{d1} \leq 1/4$ the only independent parameter or $P_{d1}=0$ and either $P_{d2R}$ or $P_{d2L}$ zero, which is equivalent to what we call the zigzag regime below. Hence, we consider the former solution. In this case, the secular equations yields
\begin{equation} \label{eq:secular1}
1 - \lambda^6 + \lambda^2 \left[ \lambda^2 h(\bm k) - h(\bm k)^* \right] \left( 1 - 4 P_{d1} \sin^2 \phi \right) + 2i \lambda^3 ( 2 \sqrt{P_{d1}} \sin \phi )^3=0,
\end{equation}
with $\lambda = e^{i(El/\hbar v+\varphi)}$ and $h(\bm k)=e^{ik_1}+e^{ik_2}+e^{-i(k_1+k_2)}$. If we define $\sin \phi' = 2\sqrt{P_{d1}} \sin \phi$, which always has a solution for $\phi'$ since $0<P_{d1} \leq 1/4$, we can write the secular equation as 
\begin{equation} \label{eq:secular2}
1 - \lambda^6 + \lambda^2 \left[ \lambda^2 h(\bm k) - h(\bm k)^* \right] \cos^2 \phi' + 2i \lambda^3 \sin^3 \phi'=0,
\end{equation}
which is equivalent to the case $P_{d1}=P_{d2R}=P_{d2L}=1/4$ and $\phi \rightarrow \phi'$. Hence, it is reasonable to assume that $\mathcal S(\phi,P_{d1})$ is unitary equivalent to $\mathcal S(\phi',1/4)$. The latter $S$-matrix is given in Eq.\ ({\color{red} 1}) of the main text where we drop the prime on $\phi$ from now on. For general $\phi$, Eq.\ \eqref{eq:secular2} has analytical solutions only at $\bar \Gamma$, in which case $h=3$ and we find
\begin{equation}
\lambda_{1\pm} = \pm e^{\mp i \phi}, \qquad \lambda_{2\pm} = \pm \exp \left[ \pm i\arctan \left( \frac{\sin \phi}{\sqrt{4-\sin^2\phi}} \right) \right],
\end{equation}
where the $\lambda_{2\pm}$ are doubly degenerate. Hence, for each network energy period $2\pi \hbar v/l$, there are always two protected nodes at the $\bar \Gamma$ point. This is shown in Fig.\ \ref{fig:sfig1} where we show the network spectrum along high-symmetry lines of the moir\'e Brillouin zone (MBZ) for several values of $\phi$. The same statement also holds at $\bar K$ and $\bar K'$. Furthermore, when $\phi\neq n\pi$ ($n \in \mathbb Z$), we find that the triple degeneracy at the $\bar \Gamma$, $\bar K$, and $\bar K'$ points of the MBZ is lifted, while the bands remain doubly degenerate at these points for all $\phi$, even after including forward scattering, as these crossings are protected by $C_3$ and $C_2 T$ symmetry.

We have shown that up to a unitary transformation, the most general $S$-matrix in the absence of forward scattering is given by $\mathcal S(\phi,1/4)$ for which the left and right interchannel deflection amplitudes are equal. Hence we consider this case from now on and perform another unitary transformation
\begin{equation}
U^\dag \mathcal S U = \frac{e^{i\varphi}}{2} \begin{pmatrix}
i \sin \phi \, S^t & e^{-i\phi} \cos \phi \, S \\
e^{i\phi} \cos \phi \, S^t & i \sin \phi \, S
\end{pmatrix}, \qquad S = \begin{pmatrix} 0 & 1 & 0 \\ 0 & 0 & 1 \\ 1 & 0 & 0 \end{pmatrix},
\end{equation}
where $U = \left[ \mathds 1_6 + i \sigma_y e^{i\phi\sigma_z} \otimes \mathds 1_3 \right]/\sqrt{2}$ transforms $a,a' \rightarrow a_\pm= \left( a\pm a' e^{\mp i\phi} \right) / \sqrt{2}$ and similar for outgoing modes. We see that for $\phi=n \pi$ ($n \in \mathbb Z$),
\begin{equation}
U^\dag \mathcal S U = \frac{e^{i\varphi}}{2} \begin{pmatrix} 0 & S \\ S^t & 0 \end{pmatrix},
\end{equation}
such that scattering modes form three independent chiral zigzag channels. Furthermore, in this case we see from Eqs.\ \eqref{eq:secular1} and \eqref{eq:secular2} that $\phi'=\phi$ such that the network supports chiral zigzag modes for any allowed values of the deflection probabilities. On the other hand, for $\phi = (n+1/2) \pi$ ($n \in \mathbb Z$),
\begin{equation}
U^\dag \mathcal S U = (-1)^n \frac{ie^{i\varphi}}{2} \begin{pmatrix} S^t & 0 \\ 0 & S \end{pmatrix},
\end{equation}
such that scattering modes perform closed orbits around AB and BA domains.

\section{S3.\quad S-matrix with forward scattering}

\begin{figure}
\centering
\includegraphics[width=.3\linewidth]{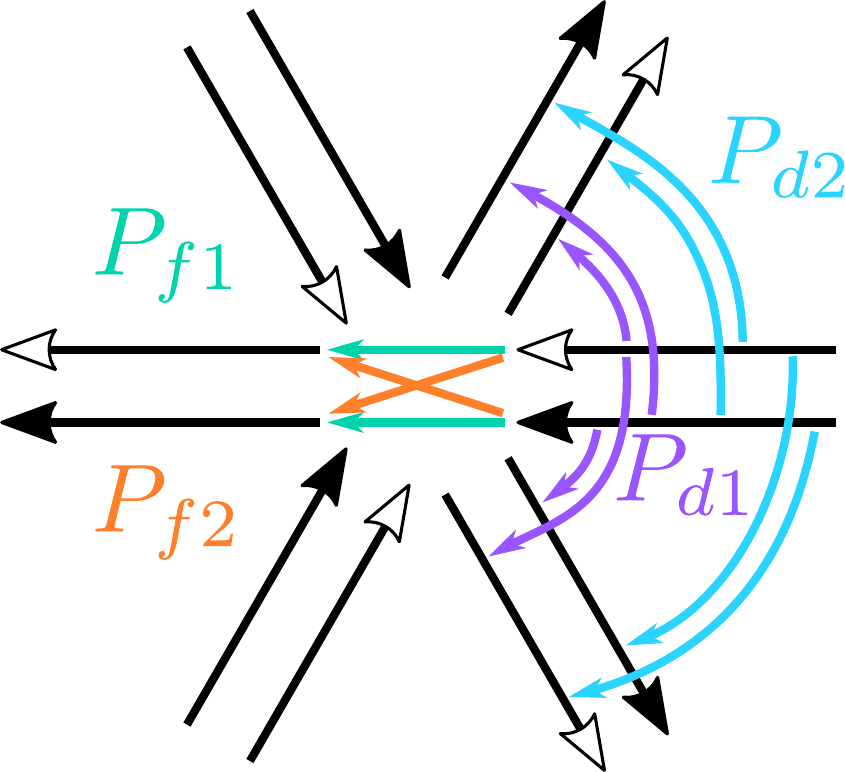}
\caption{Scattering processes for the $S$-matrix in Eq.\ \eqref{eq:Smat2} and their probabilities.}
\label{fig:sfig2}
\end{figure}
\begin{figure}
\centering
\includegraphics[width=.6\linewidth]{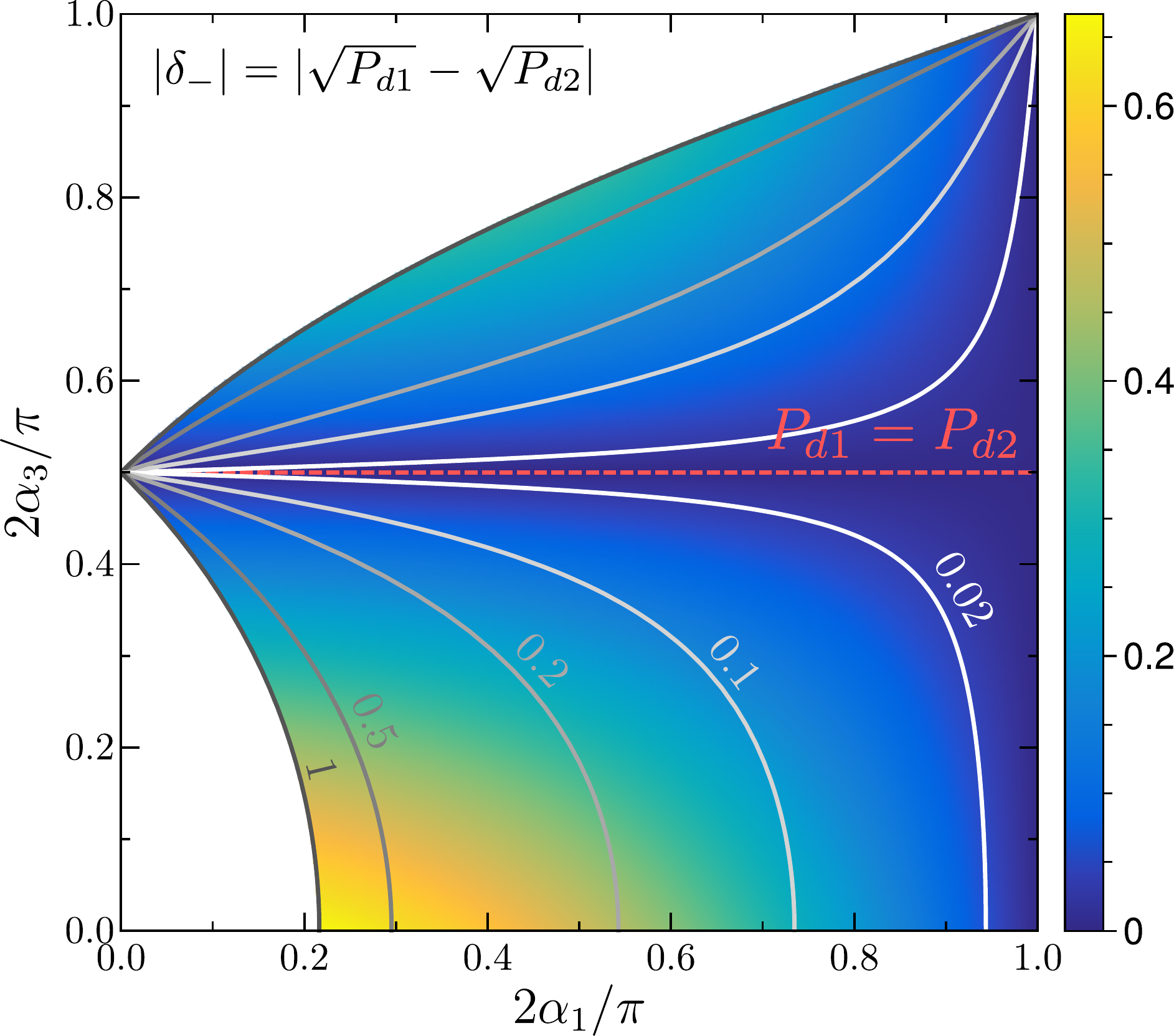}
\caption{(color online) Scattering amplitude $|\delta_-| = |\sqrt{P_{d1}} - \sqrt{P_{d2}}|$ between different zigzag modes for $\phi=0$, in the $(\alpha_1,\alpha_3)$-plane where white regions are incompatible with unitarity for any $\alpha_2$. The area enclosed by the gray-scaled curves and the right-vertical axis correspond to allowed $(\alpha_1,\alpha_3)$ for given $2\alpha_2/\pi$ as denoted next to the curves.}
\label{fig:sfig3}
\end{figure}

When we allow for forward scattering, the $S$-matrix can be written as
\begin{equation} \label{eq:Smat2}
\mathcal S =
e^{i\varphi} \begin{pmatrix}
e^{i(\phi+\chi)} \sqrt{P_{f1}} & e^{i\phi} \sqrt{P_{d1}} & e^{i\phi} \sqrt{P_{d1}} & -\sqrt{P_{f2}} & \sqrt{P_{d2}} & - \sqrt{P_{d2}} \\
e^{i\phi} \sqrt{P_{d1}} & e^{i(\phi+\chi)} \sqrt{P_{f1}} & e^{i\phi} \sqrt{P_{d1}} & -\sqrt{P_{d2}} & -\sqrt{P_{f2}} & \sqrt{P_{d2}} \\
e^{i\phi} \sqrt{P_{d1}} & e^{i\phi} \sqrt{P_{d1}} & e^{i(\phi+\chi)} \sqrt{P_{f1}} & \sqrt{P_{d2}} & - \sqrt{P_{d2}} & -\sqrt{P_{f2}} \\
-\sqrt{P_{f2}} & -\sqrt{P_{d2}} & \sqrt{P_{d2}} & -e^{-i(\phi+\chi)} \sqrt{P_{f1}} & -e^{-i\phi} \sqrt{P_{d1}} & -e^{-i\phi} \sqrt{P_{d1}} \\
\sqrt{P_{d2}} & -\sqrt{P_{f2}} & -\sqrt{P_{d2}} & -e^{-i\phi} \sqrt{P_{d1}} & -e^{-i(\phi+\chi)} \sqrt{P_{f1}} & -e^{-i\phi} \sqrt{P_{d1}} \\
-\sqrt{P_{d2}} & \sqrt{P_{d2}} & -\sqrt{P_{f2}} & -e^{-i\phi} \sqrt{P_{d1}} & -e^{-i\phi} \sqrt{P_{d1}} & -e^{-i(\phi+\chi)} \sqrt{P_{f1}}
\end{pmatrix},
\end{equation}
with $\cos\chi = \left( P_{d2} - P_{d1} \right) / 2 \sqrt{P_{f1} P_{d1}}$ such that $\chi$ is real, so that $2 \sqrt{P_{f1} P_{d1}} \geq | P_{d2}-P_{d1} |$. Note that when $P_{d2}=0$, this condition gives a lower bound on forward scattering $P_{f1} \geq P_{d1}/4$. Here, we assumed that the probability for intrachannel processes is the same for the two valley Hall states. Current conservation then requires $2(P_{d1} + P_{d2}) + P_{f1} + P_{f2} = 1$, where $P_{f1}$ ($P_{d1}$) and $P_{f2}$ ($P_{d2}$) are the probabilities for intra- and interchannel forward scattering (deflections), respectively, as illustrated in Fig.\ \ref{fig:sfig2}. We take the parameterization
\begin{equation}
P_{f1} = (\sin \alpha_1 \sin \alpha_2)^2, \quad P_{f2} = (\sin \alpha_1 \cos \alpha_2)^2, \quad P_{d1} = \frac{1}{2} (\cos \alpha_1 \cos \alpha_3)^2, \quad P_{d2} = \frac{1}{2} (\cos \alpha_1 \sin \alpha_3)^2,
\end{equation}
with $\alpha_{1,2,3} \in [0,\pi/2]$ under the condition that $\chi$ is real. This condition is graphically represented in Fig.\ \ref{fig:sfig3} where we show $|\delta_- = | \sqrt{P_{d1}} - \sqrt{P_{d2}} |$ for allowed $(\alpha_1,\alpha_3)$ and $\alpha_2=\pi/2$ ($P_{f2}=0$). Reducing $\alpha_2$ shrinks the allowed area, which in the figure corresponds to the area enclosed by the gray-scaled curves and the right-vertical axis.

\begin{figure}
\centering
\includegraphics[width=\linewidth]{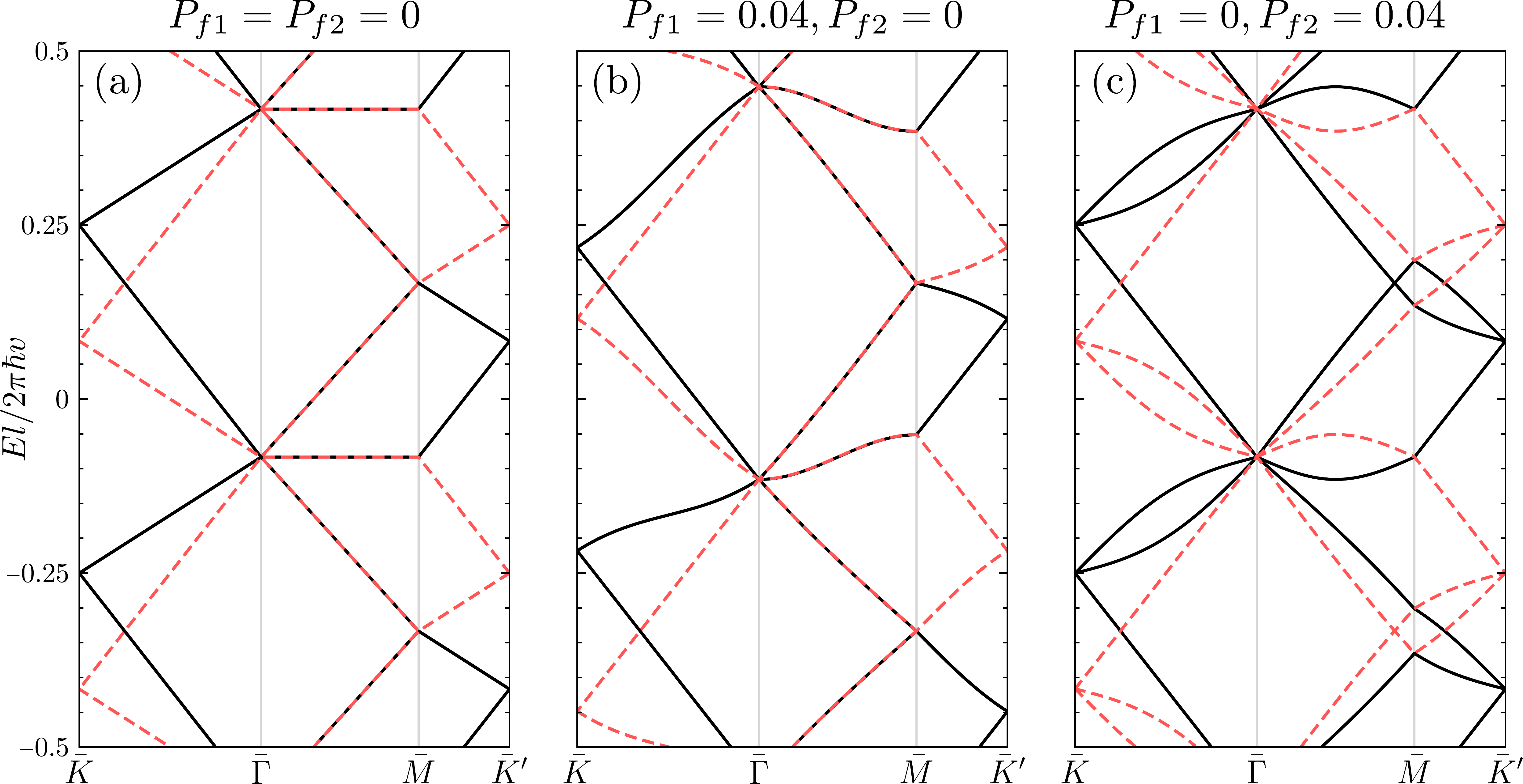}
\caption{Network spectrum in the absence of coupling between different zigzag modes ($\phi=0$ and $P_{d1}=P_{d2}$) over one energy period $2\pi \hbar v/l$ along high-symmetry lines of the moir\'e Brillouin zone as shown in the inset of Fig.\ {\color{red} 5}(a) with $\varphi =-\pi \hbar v/6l$. The bands are shown for $K$ (solid) and $K'$ (dashed) for (a) no forward scattering [Fig.\ {\color{red} 1}(c) of the main text], (b) $P_{f1}=0.02$ and $P_{f2}=0$ [Fig.\ {\color{red} 3}(b)  of the main text], and (c) $P_{f1}=0$ and $P_{f2}=0.02$ [Fig.\ {\color{red} 3}(c) of the main text].}
\label{fig:sfig4}
\end{figure}

When chiral zigzag modes propagating along different directions remain decoupled ($\phi=0$ and $P_{d1} = P_{d2}$), the Fermi surface is always nested and the density of states is constant, regardless of forward scattering, as demonstrated in the main text. We show the network bands along high-symmetry lines of the MBZ in Fig.\ \ref{fig:sfig4} for the case without and with forward scattering. On the other hand, when zigzag modes propagating along different directions are coupled ($P_{d1} \neq P_{d2}$ or $\phi\neq0$) the network bands develop anti-crossings except at the $\bar \Gamma$, $\bar K$, and $\bar K'$ points of the MBZ, where crossings are protected by $C_3$ and $C_2T$, as shown in Fig.\ {\color{red} 5} of the main text.

\end{document}